# Multi Stage based Time Series Analysis of User Activity on Touch Sensitive Surfaces in Highly Noise Susceptible Environments

Sandeep Vanga
System LSI Group, Samsung R&D Institute,
Bangalore, India

Sachin Jaganade
System LSI Group, Samsung R&D Institute,
Bangalore, India

## ABSTRACT
This article proposes a multistage framework for time series analysis of user activity on touch sensitive surfaces in noisy environments. Here multiple methods are put together in multi stage framework; including moving average, moving median, linear regression, kernel density estimation, partial differential equations and Kalman filter. The proposed three stage filter consisting of partial differential equation based denoising, Kalman filter and moving average method provides ~25% better noise reduction than other methods according to Mean Squared Error (MSE) criterion in highly noise susceptible environments. Apart from synthetic data, we also obtained real world data like hand writing, finger/stylus drags etc. on touch screens in the presence of high noise such as unauthorized charger noise or display noise and validated our algorithms. Furthermore, the proposed algorithm performs qualitatively better than the existing solutions for touch panels of the high end hand held devices available in the consumer electronics market qualitatively.

## 1. INTRODUCTION
Touch sensors have become ubiquitous and their applications span from mobile phones, personal digital assistants (PDA) to home appliances and industrial automation. Touch sensors are signal transducers, which convert one form of energy to other, that are sensitive to touch. This can be realized using many techniques such as capacitive, resistive, infrared, optical, Surface Acoustic Waves (SAW) etc [1-4].

Resistive touch sensors consist two layers of electrodes which are not in touch with each other. When an external object touches the surface of sensor, it forces these two layers to come in to contact and triggers a flow of current which can be sensed easily. Though cheaper, as resistive touch sensors depend on the amount of pressure, their sensitivity is not so great [5]. Capacitive sensing is one of the most promising touch sensing technologies and is widely used in smart phones, tablets and surfaces due to its capability to detect multi touch, sensitivity to respond for human touch. Also, compared to resistive touch technology, capacitive touch screens can be designed to be much thinner [6-7]. Over the past decade with the advent of iPhone, Samsung Galaxy devices, capacitive touch sensing has emerged as the most prominent touch sensing technology.

Infrared or optical sensors keep emitting light and when an object comes in to the vicinity of the touch surface these beams are obstructed and can be detected [8]. SAW sensors form grid ultrasonic waves across the touch surface and any activity near to the touch surface is identified very easily as the grid is disturbed [9]. Usually Infrared and SAW touch sensors are employed in larger displays whereas capacitive touch panels are being used widely in smaller displays such as smart phones.

There are many variants of capacitive touch technology and the most popular and extensively used variant is called Projected Capacitive Touch (PCT) technology. PCT can be implemented in two different ways such as Self and mutual capacitance. Self capacitance model measures the capacitance of electrode with respect to ground where as mutual capacitance measures the capacitance between the two conductors overlaid on top of each other. In case of self capacitance when a conducting object (finger/stylus) is brought closer to the touch surface, it forms a capacitance with respect to the electrode and the self capacitance is changed. In mutual capacitance model, conducting object (finger/stylus) steals the charge flowing between the two electrodes and hence the change in capacitance [10-11].

Typically these electrodes are arranged in two layers to form rows and columns and at each intersection mutual capacitance is measured. This enables the high resolution and multi touch capability. In case of self capacitance approach each electrode is treated independently and the self capacitance is measured for each row and column, due to which multi touch is impossible. Given 'M' electrodes in upper layer and 'N' electrodes in the bottom layer; in mutual capacitance setup 'MxN' values are measured and in self capacitance set up 'M+N' values are measured [10-13].

There are numerous advantages of PCT technology such as high longevity, transmittance, reliability, sharp response and multi touch capability (in case of mutual capacitance only). One of the major disadvantage of PCT technology is that its sensitivity to Electro Magnetic Interference (EMI) and there are many such sources which cause EMI such as Liquid Crystal Display (LCD), unauthorized chargers etc [10-11]. So there is an immense need to take care of disturbances caused due to such noise sources. These noise sources generate high amounts of noise affect user activity (finger/stylus drags) adversely. This can be handled either at hardware level or at firmware (software) level [14-15].

Typical software architecture for touch handling touch sensor data consists of pre emphasis, segment identification and rejection, coordinate extraction, object (finger/stylus) tracking, activity (gestures) recognition etc [1,16-20]. This paper discusses one such software approach to reduce the effect of noise by applying smoothing filters on the output of coordinate extraction block based on time series analysis.

Time series analysis has been pursued as active area of research spanning across multiple disciplines such as econometrics, signal processing, machine learning, computational finance, weather forecasting, seismology, statistics, geophysics etc. Several approaches were proposed in the area of time series analysis such as auto regressive models, wavelet based methods, linear and non linear regression, partial differential equations (PDE) etc [21-23].

Further we can categorize time series analysis problem in to different sub problems such as prediction or forecasting task, smoothing or noise removal task, signal estimation task based on particular area of interest [22-23]. Here we are addressing the problem of predicting signal using current and previous inputs in highly noise susceptible environments. A novel multi stage framework using feedback loops is proposed to combine multiple methods to solve the above mentioned problem. This frame work outperforms existing methods. Quantitative performance measures like Mean Squared Error (MSE), Maximum Error are used to evaluate the proposed algorithms.

## 2. NOISE SMOOTHENING TECHNIQUES
We experimented with multiple methods including moving average, moving median, kernel density estimators, linear regression, partial differential equations and Kalman filter for noise reduction and prediction tasks.

### 2.1 Modified Moving Average/Median Filter
In case of normal moving average/median filter, current value is replaced by average/median of previous 'n' values, current value and future 'n' values, which in turn causes a group delay of 'n' [22-23]. Here, in the modified version previous 'n' values from the output of the modified moving average filter are used instead of previous 'n' values. The idea behind this change is to feed best possible values available at time as input to modified moving average/median filter. It is obvious that the modified version also causes frame delay of 'n'.

Similarly, normal median filter and modified median also can be defined by replacing average operator with median operator in the above equations. In the context of finger/stylus drag smoothing, each input is a two dimensional vector consisting of X and Y coordinates on the touch surface.

### 2.2 Odd One Removed Moving Average Filter and Variants
A variant of moving average filter is also proposed, which uses only '2n' values instead of '2n+1' values by eliminating the most dissimilar one which can be called as "odd one". Hence the name "odd one removed out moving average filter". This is done by calculating the distance from a reference point to all values and removing the one which is most distant from the reference point. Here, we propose four versions of such filters and these filters mainly differ based on calculation of the reference point.

(A) In first variant, we use previous smoothed output of the filter as reference point

(B) In second variant, we use average of; previous 'n' smoothed outputs of the filter, current and future 'n' values of noisy data as reference point

(C) In third variant, we use previous smoothed output of the filter as reference point. But, distance from reference point to each coordinate is calculated independently for X, Y coordinates.

(D) In fourth variant we use same reference point as second variant, but distance is calculated independently for X, Y coordinates.

Third and fourth variants cannot be defined as independent smoothing filters, but they are used in conjunction with linear regression filter to eliminate odd dependant variable. Linear regression filters will be explained in detail further in the paper.

### 2.3 Kernel Density Estimation (KDE) based Filter
The KDE based approach used here is similar to the approach used for image smoothing technique [24]. This approach is motivated from the mean shift technique [25]. This technique is similar to weighted averaging where weights are selected using Gaussian Kernel. These weights are inversely proportional to the distance of the data from the current data point.

This weighted average is repeated iteratively until it converges based on stopping criteria. The old noisy data is replaced by new smoothed data points from iteration to iteration. The stopping criteria used here is based on the change in Standard Deviation (SD) of data or change in point to point distance. If there is no change in SD or updated data then it is assumed to be converged.

The KDE filter takes '2n+1' data points as input i.e. previous 'n', future 'n' and the current noisy value. It is applied to x and y co-ordinates separately.

Smoothed outputs are calculated for all data points in a given iteration. This iterative process is repeated until convergence as explained earlier.

Smoothed outputs of Y coordinates also can be calculated using KDE filter in similar fashion.

### 2.4 Linear Regression based Filter
Linear regression is well known approach [22] used for modeling a dependant variable based on one or more independent variables. In the context of touch sensory signals, time (frames) is independent variable and X, Y coordinates are dependant variables. Here, we try to derive relationship between time and X coordinates, time and Y coordinates separately using linear regression methods to model variation of X, Y with respect to time.

As we have only one independent variable, simple linear regression model can be used [26]. There, are many approaches to solve linear regression. Some of them are least squares approach [22], Theil-sen approach [27] etc. We experimented with all of these approaches and using least squares linear regression model is best suited for touch sensor signals.

This filter performs linear regression on a set of values by optimizing sum of squares cost function. In other words it tries to fit a line through given set of points using least squares criteria. Here, we are using moving window based line fit with 'm' previous inputs and 'n' future inputs. So, group delay of 'n' is added in by this filter.

As time is independent variable and X, Y coordinates are dependant variables, regression line is fit for X, Y coordinates independently as a function of time. Here, we also use third or fourth variant of odd one removed out filter to eliminate one such odd value before performing regression. This is done on both X, Y coordinates separately.

## 2.5 Smoothing in Polar Coordinate System

Each point on touch surface is consisting of X and Y coordinates. We convert it to polar coordinates by shifting origin to first coordinate of the drag, so each $(X, Y)$ is converted to $(R, \theta)$. Once polar coordinates are derived, we use modified moving average filter to suppress noisy R. Theta ($\theta$) is constant, given underlying signal is varying linearly with time. So, we use first order exponential smoothing [22, 28] on top of modified moving average filter for smoothening out Theta.

## 2.6 Kalman Filter

We used Kalman filter [22, 29-30] to suppress noise by assuming a pre-defined constant acceleration model of a system. It consists of two steps a) Prediction b) Correction. Using dynamic model, state is predicted in prediction state while the state is corrected based on observation model. Kalman does not cause in any group delay

Basic components of Kalman filters are State vector, Dynamic model and Observation model. Each component is described in detail below.

### 2.6.1 State Vector

State vector describes the state of dynamic system. As Kalman filter consists of two steps, state vector has two values, one in predication state and other after correction state. Here, state consists of signal value at a time point, velocity of the signal as well as the acceleration of the signal.

### 2.6.2 Dynamic Model

Dynamic model denotes the transformation of the state vector over time. In the linear case this can be given as below.

$$X[k] = A * X[k-1] + q; \quad q \approx N(0, Q)$$

Where A is the state transion matrix and is constant $X[k]$ is the state vector and q is the dynamic noise which is usually assumed as white noise. In our present model we have used $X = (x, y, x', y', x'', y'')^T$; where $(x', y')$ represents velocity, $(x'', y'')$ represents acceleration.

### 2.6.3 Observation Model

The observation model defines the relationship between the state and the measurements. The measurements can be described by a system of linear equations in case of linear model, which depends on the state variables. The matrix form of this system is

$$Y[k] = H * X[k] + r; \quad r \approx N(0, R)$$

Where $Y[k]$ is the observation vector, H is the state transition matrix and is constant, r is the noise of the measurement process with the covariance matrix R. In our model state to observation transition matrix is given by,

$$H = \begin{bmatrix} 1 & 0 & 0 & 0 & 0 & 0 \\ 0 & 1 & 0 & 0 & 0 & 0 \end{bmatrix}$$

### 2.6.4 Noise Modeling

We calculate process noise matrix and measurement noise matrix as follows. First noise matrix Q is calculated. Standard Deviation (SD) of previous 'n' points is calculated for x and y coordinates seperately. Say SD of X, Y cordinates is sd_x and sd_y respectively.

$$sd\_const = (sd\_x + sd\_y)/2$$

Define $\quad q = 0.01 * sd\_const$

Process noise matrix Q is calculated as follows,

$$Q = q * Q_c$$

and $Q_c$ is given by discretization of the continous-time system. For state transition matrix A we have used MatLab function lti_disc.

Measurement noise matrix R is given by,

$$R = \begin{bmatrix} sd\_x & 0 \\ 0 & sd\_y \end{bmatrix}$$

Process error covariance matrix P is calculated as follows. Initialize P as given below.

$$P = SCALE\_FACT * Q; \quad SCALE\_FACT \text{ is a scalar.}$$

For each point, we use previous noisy points to calculate sd_x and sd_y and then update Q and R before passing it for Kalman filter. Kalman filter will update and correct P. Predict and update stages are defined by system equations [29-30].

## 2.7 Partial Differential Equation (PDE) based Smoothing

Partial differential equations (PDEs) are used to describe wide variety of phenomenon in the real world like heat flow, fluid flow etc. PDEs are equations that involve rates of change of a desired quantity with respect to the underlying independent variables in the system [31]. In recent times, it's used for image enhancement also. Below mentioned heat equation is used in image noise reduction [32-33].

$$\partial I(x, y, t) / \partial t = \nabla.(c(x, y, t)) \nabla I(x, y, t)$$

Where, $I(x, y, t)$ is noisy image and $c(x, y, t)$ is influence coefficient. Here, we assume that noisy image is similar to a 2D surface heated up unevenly at particular locations. So, we use PDE, which describes the heat flow from high temperature regions to low temperature region over the time, to suppress noise in the image. Similarly, we can apply the same technique to reduce noise in any given time series [34]. Here we can assume that noisy finger drag is similar to a thin wire which is heated up unevenly at particular locations. Below mentioned equation can be used to get the smoothed output of the noisy finger drag.

$$S(x, t + \Delta t) = S(x, t) + \Delta t(d_f c_f + d_b c_b)$$

$$d_f = S(x - \Delta x, t) - S(x, t) \quad c_f = \frac{1}{1 + (d_f / k)^2}$$

$$d_b = S(x + \Delta x, t) - S(x, t) \quad c_b = \frac{1}{1 + (d_b / k)^2}$$

Here, $S(x, t)$ is one dimensional noisy signal, $d_f$ and $d_b$ are gradients in forward and backward directions, $c_f$ and $c_b$ are corresponding influence coefficients respectively. $k$ is a constant which influences contribution of the gradients to the noise reduction step. $\Delta t$ is the step size of the noise reduction for each iteration. Higher $\Delta t$ values help to converge faster but, noise reduction might not be effective. Whereas for lower $\Delta t$ values reduce noise very effectively, but converge very slowly. So we need to select moderate step size in order to strike balance between convergence and noise reduction. $\Delta x$ is the sampling rate of the noisy signal. In case of finger drag, we treat each of the X-coordinates and Y-coordinates as independent one dimentional noisy signals. We use above mentioned smoothing technique to reduce the jitter in input signal. Here, we use $k = 100$ and $\Delta t = 0.25$. We repeat the same steps for 1000 iterations or till it converges.

Let's assume $x(t-1)$, $x(t)$, and $x(t+1)$ are jittery data inputs to the filter, after applying PDE based smoothing we get smoothed estimate for $x(t)$ as $O_s(t)$. Now we continue smoothing using the refined input $O_s(t).x(t+1)x(t+2)$ to get smoothed estimate for $x(t+1)$ and so on. Here, as we need only one future value as input to the filter, PDE based smoothing always introduces one frame delay only.

## 3. MULTI STAGE FRAMEWORK

Though the filters explained in Section 2 perform well in prediction and noise reduction tasks, combining them in a novel framework gives much better results than using them standalone. Here a multi stage frame work is proposed to combine some of these techniques in a novel way. Main strength of the multi stage frame work lies in the fact that it uses feed forward and feedback loops to pass smoothed signal from one filter to other filter.

We devised such multiple filters using different combinations of predictive or noise reduction techniques mentioned in the previous section. Below are some of typical examples of the filters we explored.

(a) Moving median filter output feed forwarded to odd one removed out moving average filter whose output is feed backed to moving median filter

(b) Kernel density estimator output is feed forwarded to moving average filter whose output is feed backed to kernel density estimator

(c) Odd one removed out moving average filter output feed forwarded to least squares linear regression filter whose output is feed forwarded to moving median filter whose output is feed backed to odd one removed out moving average filter

(d) Partial differential smoothing output feed forwarded to Kalman filter whose output is feed forwarded to moving average filter and Kalman filter output is feed backed to PDE based smoothing filter

We also experimented with replacing moving average filter by Savitzky-Golay filter.

### 3.1 Savitzky-Golay (SG) Filter

This filter performs polynomial regression on set of values. It is a weighted average filter and weights change as we change number of points and polynomial order [35]. Here, we typically use polynomial order of two and five points (taps). This method preserves some important features like relative maxima, minima, width of the peaks etc. So definitely, this is not a good choice for first stage. But it improves the performance of first stage filters when used in second stage. Because of these properties this method suppresses noise while preserving the genuine signal changes.

It can be used on top of any filter output to get slight improvement. We applied it on top of modified moving averae, modified running median and first variant of odd one removed moving average filters. In case of second order polynomial with five tap SG filter, weights are given as, [-0.086 0.343 0.486 0.343 -0.086]. A typical two stage filter using SG in second stage is given as below.

## 4. RESULTS AND DISCUSSION

We generated time series synthetically by adding high amounts of noise both along the direction of movement of the time series as well as perpendicular to the direction of the movement. We also generated both linear and non linear trajectories of time series. We also tested our data on time series with varying velocity and accelerations. Finally, we obtained real world data like hand writing, finger/stylus drags etc. in the presence of high amounts of charger/display noise from touch interfaces of hand held devices and validated our algorithms.

We used both quantitative as well as qualitative measures to compare performance of different algorithms. Some of the Quantitative measures include Average Euclidean distance or Mean Squared Error (MSE) between actual time series and smoothed time series and SNR improvement etc. Among all the filters we experimented with, proposed three stage Filter 'd' mentioned in the earlier section (PDE + KF + Moving Average) gives best results.

To compare performance of the proposed three stage filter we passed the same input (noisy time series) through different multi stage filters mentioned earlier. We made sure that group delay of all the filters is same. Below we can see perceptual performance difference between "Proposed Three Stage" filter and "Modified Moving Average" filter with five frame group delays.

Though Modified Moving Average Filter performs decently it is definitely no match to the Proposed Three Stage filter's performance. Even in case of non linear drags, Proposed Three Stage filter is superior to Modified Moving Average filter. Below we can see perceptual performance difference between proposed three stage filter and Modified Moving Average filter with five frame group delay.

Also, quantitative results are tabulated for various types of drags shown above (Linear, Non Linear and Zigzag Drags) with different velocity and acceleration combinations. Each drag is generated synthetically 100 times and the quantitative measures are calculated across these 100 samples. Normal distributed random noise is added in both the same direction of the drag as well as the perpendicular direction of the drag. As the noise characteristics observed on various touch panels have more noise in the perpendicular direction of the drag compared with the noise same direction as the drag, noise simulations are modeled to reproduce such noise conditions as closely as possible.

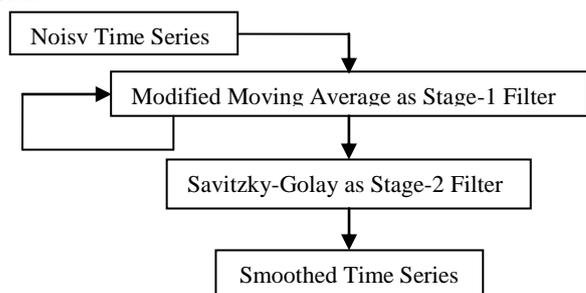

**Figure 1: A Two Stage Filter with SG filter on top of Modified Moving Average Filter**

A typical example of three stage filter is given as below.

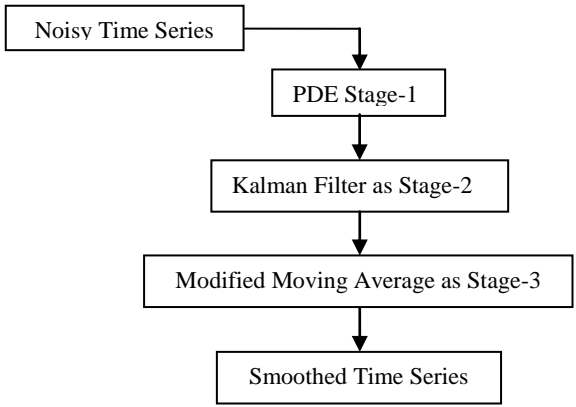

**Figure 2: A typical three stage filter proposed for time series analysis**

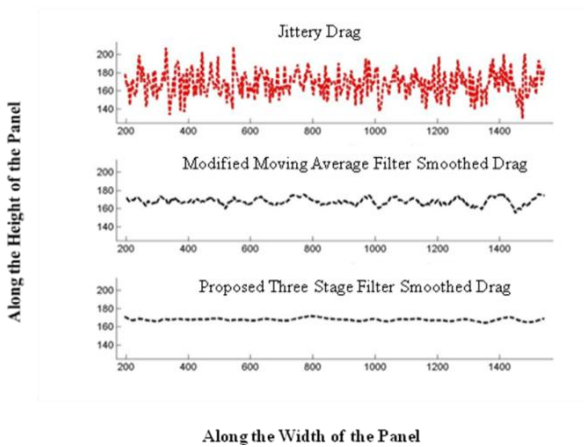

**Figure 3: Smoothed time series versus with Noisy time series of Linear Drag**

Let's say original drag (signal/time series) is defined as $S = (S_1, S_2, ..., S_M)$ and the noisy data generated on top of original time series S is given by $N = (n_1, n_2, ..., n_M)$ and corresponding filtered output is given by $F = (f_1, f_2, ..., f_M)$ . As explained earlier 100 different noisy time series $N^1, N^2, N^3, ..., N^{100}$ are synthetically generated for each drag S and corresponding filtered outputs are defined by $F^1, F^2, F^3, ..., F^{100}$.

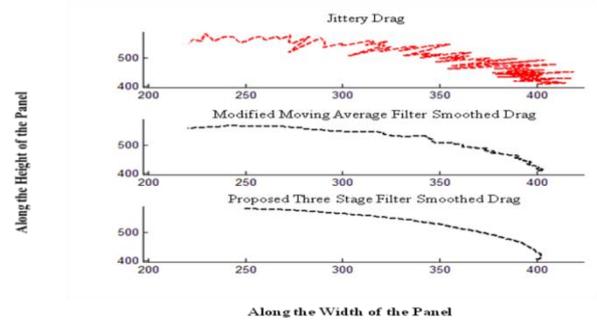

**Figure 4: Smoothed time series versus with Noisy time series of Non Linear Drag**

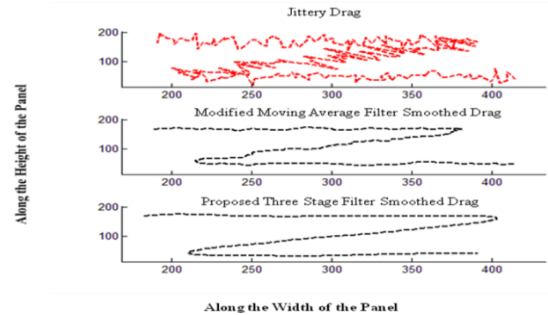

**Figure 5: Smoothed time series versus with Noisy time series of Zigzag Drag**

Here, two different quantitative measures were used as defined below to evaluate performance of the proposed filter. Modified Moving Average filter is used to compare the performance of the proposed filter. Here, as each time point consists of two coordinates, let's define $S_i$ as $S_i = (S_{xi}, S_{yi})$ . So, similarly $n_i$ and $f_i$ can also be defined as $n_i = (n_{xi}, n_{yi})$ and $f_i = (f_{xi}, f_{yi})$ respectively.

Measure1:

$$MeanSquaredError = \frac{1}{100}\sum_{j=1}^{100}\frac{\sum_{i=1}^{M}\sqrt{(s_{xi} - f_{xi})^2 + (s_{yi} - f_{yi})^2}}{M}$$

Measure2:

$$MaximumError = \max_j \max_i \sqrt{(s_{xi} - f_{xi})^2 + (s_{yi} - f_{yi})^2}$$

$where \ i = 1, ..., M; j = 1, ..., 100$

From the quantitative analysis tabulated below, it is evident that Proposed Three Stage filter performs better than Modified Moving Average filter for any given drag.

**Table 1 Quantitative measures (Means Squared Error – MSE and Maximum Error) of proposed noise suppression algorithms for synthetically generated linear drag. Each row indicates a different linear drag (shown in Figure 3) with particular velocity and acceleration combination**

| Velocity (in mm per sec) | Acceleration (in mm per sec squared) | Noisy Time Series | | Modified Moving Average Filtered Time Series | | Proposed Three Stage Filtered Time Series | |
|---|---|---|---|---|---|---|---|
| | | Mean Squared Error (in mm) | Max Error (in mm) | Mean Squared Error (in mm) | Max Error (in mm) | Mean Squared Error (in mm) | Max Error (in mm) |
| 10 | 0 | 1.34 | 5.1 | 0.41 | 1.96 | 0.29 | 1.71 |
| 25 | 0 | 1.39 | 5.2 | 0.4 | 2.1 | 0.31 | 1.44 |
| 50 | 0 | 1.35 | 5.1 | 0.4 | 1.88 | 0.32 | 1.62 |
| 100 | 0 | 1.36 | 5.2 | 0.41 | 2.2 | 0.31 | 1.73 |
| 150 | 0 | 1.35 | 5.0 | 0.4 | 1.9 | 0.33 | 1.82 |
| 200 | 0 | 1.34 | 5.1 | 0.42 | 2.1 | 0.35 | 1.94 |
| 25 | 25 | 1.35 | 5.1 | 0.4 | 2.2 | 0.29 | 1.95 |
| 25 | 50 | 1.34 | 5.2 | 0.39 | 2.1 | 0.32 | 1.82 |
| 25 | 100 | 1.37 | 5.1 | 0.41 | 1.8 | 0.31 | 1.55 |
| 100 | 25 | 1.35 | 5.0 | 0.41 | 1.9 | 0.32 | 1.94 |
| 100 | 50 | 1.34 | 5.1 | 0.39 | 2.4 | 0.3 | 2.1 |
| 100 | 100 | 1.36 | 5.2 | 0.4 | 2.1 | 0.31 | 1.89 |

**Table 2 Quantitative measures (Means Squared Error – MSE and Maximum Error) of proposed noise suppression algorithms for synthetically generated non linear drag. Each row indicates a different Non Linear drag (shown in Figure 4) with particular velocity and acceleration combination**

| Velocity (in mm per sec) | Acceleration (in mm per sec squared) | Noisy Time Series | | Modified Moving Average Filtered Time Series | | Proposed Three Stage Filtered Time Series | |
|---|---|---|---|---|---|---|---|
| | | Mean Squared Error (in mm) | Max Error (in mm) | Mean Squared Error (in mm) | Max Error (in mm) | Mean Squared Error (in mm) | Max Error (in mm) |
| 10 | 0 | 1.31 | 4.37 | 0.44 | 1.19 | 0.39 | 1.12 |
| 25 | 0 | 1.19 | 4.4 | 0.3 | 0.85 | 0.27 | 0.72 |
| 50 | 0 | 1.39 | 4.8 | 0.49 | 1.43 | 0.43 | 1.35 |
| 100 | 0 | 1.38 | 4.2 | 0.39 | 1.4 | 0.37 | 1.34 |
| 150 | 0 | 1.29 | 5.3 | 0.49 | 1.75 | 0.41 | 1.61 |
| 200 | 0 | 1.3 | 3.53 | 0.79 | 1.71 | 0.68 | 1.65 |
| 25 | 25 | 1.17 | 3.65 | 0.35 | 1.0 | 0.34 | 0.94 |
| 25 | 50 | 1.24 | 5.7 | 0.4 | 1.24 | 0.38 | 1.17 |
| 25 | 100 | 1.48 | 5.3 | 0.39 | 1.35 | 0.32 | 1.33 |
| 100 | 25 | 1.24 | 4.2 | 0.44 | 1.65 | 0.42 | 1.57 |
| 100 | 50 | 1.29 | 6.0 | 0.5 | 1.6 | 0.45 | 1.54 |
| 100 | 100 | 1.38 | 4.2 | 0.64 | 1.48 | 0.61 | 1.32 |

**Table 3 Quantitative measures (Means Squared Error – MSE and Maximum Error) of proposed noise suppression algorithms for synthetically generated Zigzag drag. Each row indicates a different Zigzag drag (shown in Figure 5) with particular velocity and acceleration combination**

| Velocity (in mm per sec) | Acceleration (in mm per sec squared) | Noisy Time Series | | Modified Moving Average Filtered Time Series | | Proposed Three Stage Filtered Time Series | |
|---|---|---|---|---|---|---|---|
| | | Mean Squared Error (in mm) | Max Error (in mm) | Mean Squared Error (in mm) | Max Error (in mm) | Mean Squared Error (in mm) | Max Error (in mm) |
| 10 | 0 | 1.29 | 3.95 | 0.36 | 1.2 | 0.34 | 1.15 |
| 25 | 0 | 1.32 | 4.5 | 0.42 | 1.62 | 0.39 | 1.46 |
| 50 | 0 | 1.37 | 4.3 | 0.45 | 1.24 | 0.41 | 1.1 |
| 100 | 0 | 1.44 | 4.6 | 0.53 | 1.89 | 0.47 | 1.71 |
| 150 | 0 | 1.33 | 4.9 | 0.52 | 2.72 | 0.47 | 2.15 |
| 200 | 0 | 1.29 | 4.1 | 0.76 | 2.89 | 0.65 | 2.03 |
| 25 | 25 | 1.34 | 4.9 | 0.35 | 1.12 | 0.28 | 1.03 |
| 25 | 50 | 1.42 | 4.8 | 0.52 | 1.7 | 0.44 | 1.62 |
| 25 | 100 | 1.17 | 4.3 | 0.38 | 1.45 | 0.28 | 1.32 |
| 100 | 25 | 1.18 | 4.65 | 0.42 | 2.1 | 0.5 | 1.87 |
| 100 | 50 | 1.24 | 4.0 | 0.53 | 2.26 | 0.47 | 2.11 |
| 100 | 100 | 1.22 | 4.15 | 0.54 | 2.47 | 0.46 | 2.21 |

## 4.1 Error Variation with Velocity and Acceleration

The figure below shows performance variation of the proposed algorithms at different velocities for synthetically generated line drags. Here, line drag is assumed to be of uniform velocity i.e. zero acceleration. Mean Squared Error (MSE) is used as performance measure.

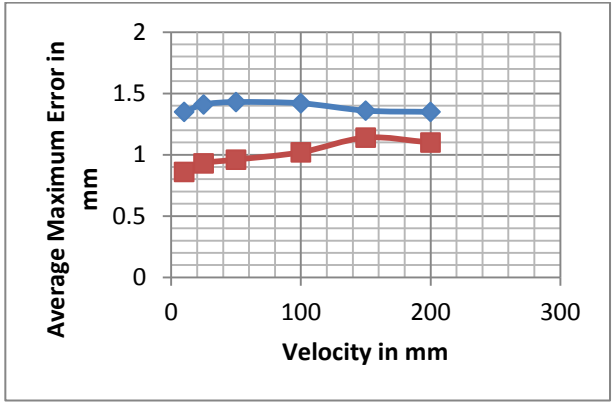

**Figure 6: Performance (Average Maximum Error) of Proposed Noise Suppression Algorithms with respect to velocity. Blue and red curves indicate the performance of Modified Moving Average and Proposed Three Stage Filters respectively.**

As the velocity increases Modified Moving Average filter performs more or less similar where as Proposed Three Stage filter's performance is degraded at higher velocities. This is due to the presence of Kalman filter in three stage filter. Kalman slowly tracks the signal when at high velocities. This indicates that three stage filter depends upon velocity. Thus velocity play important role in this case. But, notably the Proposed Three Stage filter outperforms Modified Moving Average filter at all the velocities. The figure 7 shows performance variation of the proposed algorithms for synthetically generated line drags at different accelerations. Here, we experimented with two different velocities i.e. 25mm/sec and 100mm/sec. 25mm/sec is considered to be slow drag and 100mm/sec is considered to be fast drag. Mean Squared Error (MSE) is used as performance measure.

Performance of Modified Moving Average filter is more or less same regardless of velocity and acceleration changes. In case of Proposed Three Stage filter, it performs well for slow drag compared with fast drag at all accelerations and performance is more or less same at all accelerations for fast drag. This indicates that acceleration does not affect the performance of the proposed filters as much as velocity.

## 4.2 Qualitative Performance of Proposed Algorithms on Galaxy Hand Set

Finally, we evaluated performance of proposed algorithms for finger/stylus drag using Galaxy S4 hand set in the presence of high amounts of charger/display noise. Here, both the algorithms are integrated with touch controller software suite. Time complexity of Proposed Three Stage filter is about 100 ms and Modified Moving Average is about 45 ms for a single drag at 30 MHz clock speed. This indicates that Proposed Three Stage filter is highly practical and works in real time. As shown in Figure 8 & 9, Proposed Three Stage filter's performance is much better than Modified Moving Average filter.

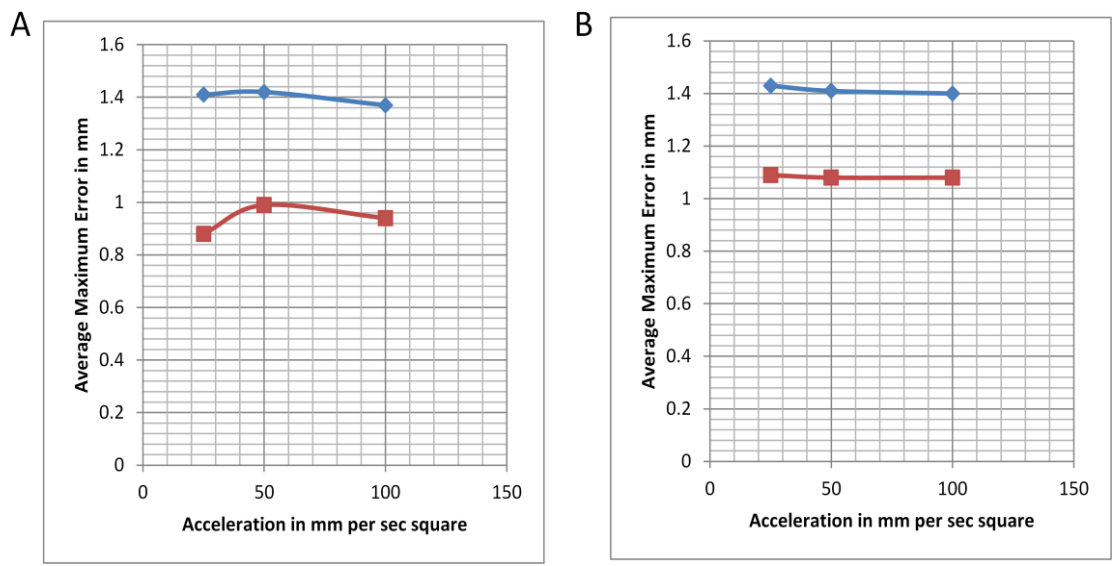

**Figure 7: Performance (Average Maximum Error) variation of Proposed Noise Suppression Algorithms as acceleration changes. Figure 7A indicates velocity at 25 mm/sec and Figure 7B indicates velocity at 100 mm/sec. Blue and red curves indicate the performance of Modified Moving Average and Proposed Three Stage Filters respectively.**

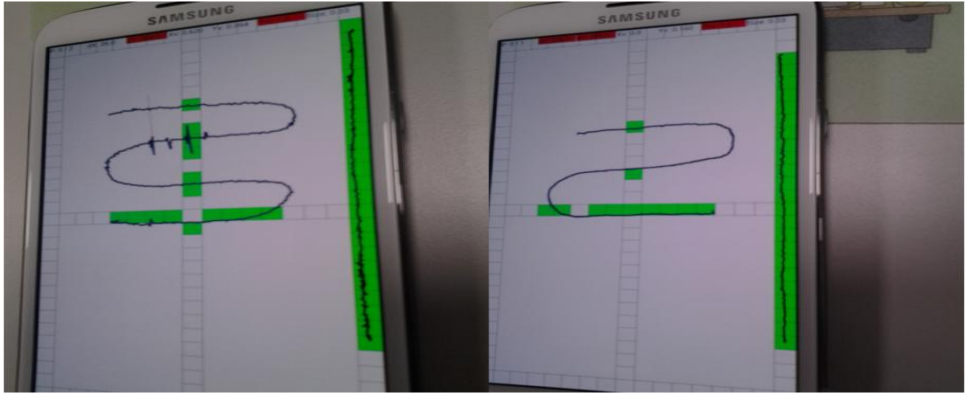

**Figure 8: Qualitative performances of proposed algorithms on Galaxy S4 Handset for finger drag in the presence of high amount of display noise. Left side of the figure depicts output of Modified Moving Average Filter where as right side of the figure depicts output of Proposed Three Stage Filter**

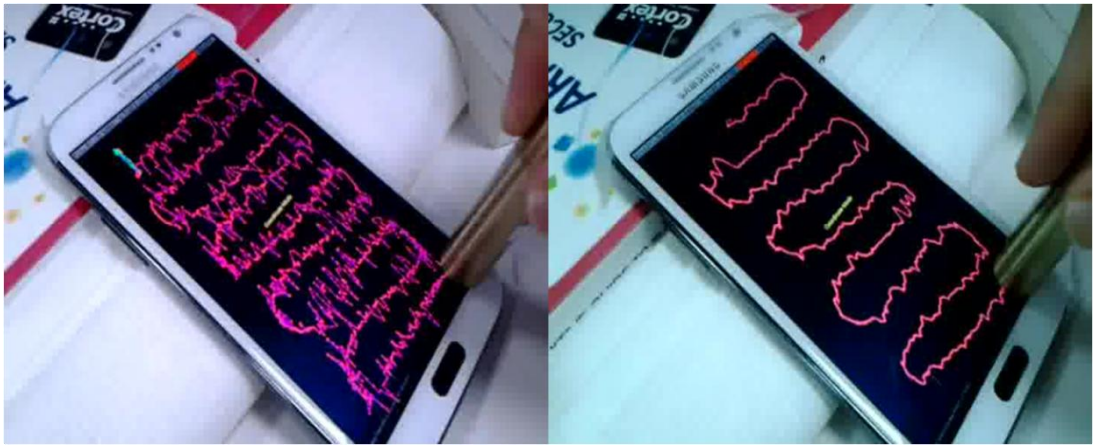

**Figure 9: Qualitative performances of proposed algorithms on Galaxy S4 Handset for finger drag in the presence of high amount of charger noise. Left side of the figure depicts output of Modified Moving Average Filter where as right side of the figure depicts output of Proposed Three Stage Filter**

## 5. CONCLUSIONS

Definitely, performance of the Proposed Three Stage filter is much better qualitatively as well as quantitatively compared with Modified Moving Average filter with slight increase in time complexity. On an average Proposed Three Stage filter gives 25% better accuracy than Modified Moving Average filter for linear as well as non linear drags. In case of highly noise susceptible environments, we observed that proposed Three Stage filter outperforms all the single stage as well as other multi stage filters proposed in this paper. In case of low noisy environments we observed that bypassing PDE gives better results. So based on the noise levels, we can adaptively switch ON/OFF PDE filter in the proposed three stage framework. Even in case of real world data captured from touch interfaces, Proposed Three Stage filter outperforms existing solutions present in highly competent products available in the consumer electronics market. The proposed idea is tested for unauthorized charger connected case, which should be tested for other types of noise sources like in presence of different light sources. Presently PDE is the part of three stage filter. The future scope for this idea is testing the same method for different noise sources and dynamically selecting PDE based on noise level.

## 6. ACKNOWLEDGEMENTS

The authors would like to thank Device Solutions Team, System LSI Division, Korea for funding the project and providing with the real world data from Galaxy handsets.